\documentclass[14pt]{article}
\usepackage[koi8-r]{inputenc}
\usepackage[russianb,english]{babel}
\usepackage{epsfig}
\usepackage{prepr}
\begin{document}

\pagestyle{prepr}
\thispagestyle{empty}
\renewcommand{\thesection}{\arabic{section}.}

\begin{center}
\epsfclipon
\epsfig{figure=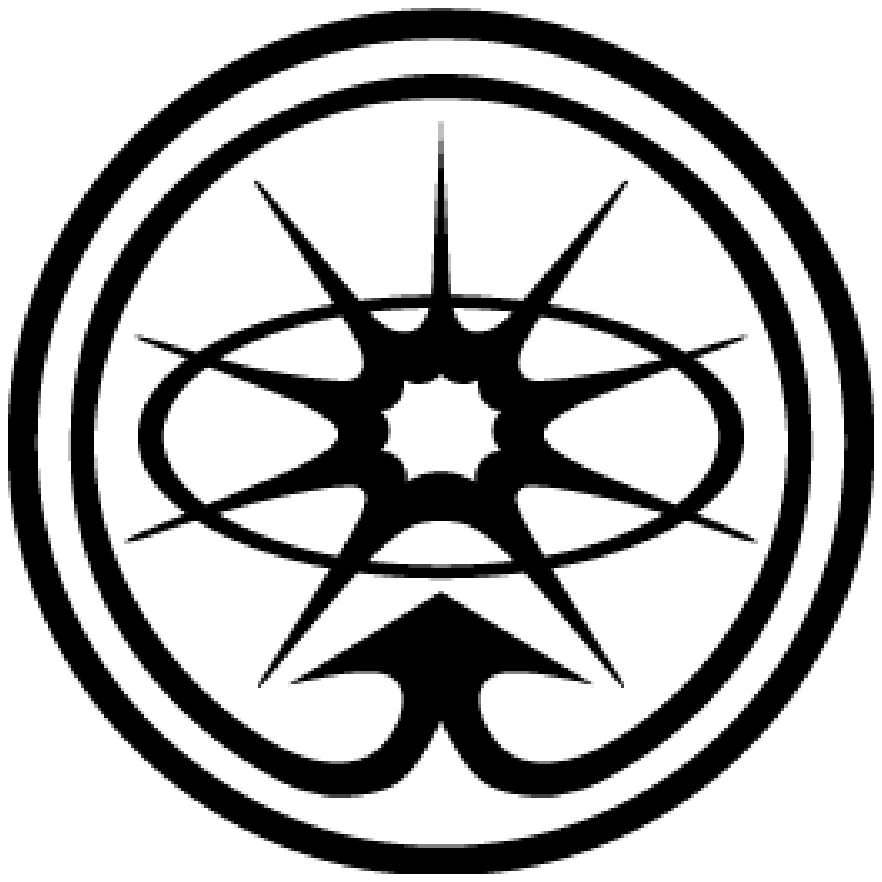, width=3cm} \hfil \parbox[b][3cm][c]{0.72\textwidth}
{\bf \LARGE \foreignlanguage{russian}{Институт \hfil Теоретической \hfil и \\ Экспериментальной \hfil Физики}} \hspace{15mm}
\vskip 1mm
{\bf \LARGE \hfill 21--06}
\vskip 30mm
\hspace{0.5cm} \parbox[b][8em][c]{0.7\textwidth}
{\sf \large 
\begin{center}
I.G.~Alekseev, V.A.~Andreev,
P.E.~Budkovsky, E.A.~Filimonov,
V.V.~Golubev, V.P.~Kanavets, M.M.~Kats,
L.I.~Koroleva, A.I.~Kovalev, N.G.~Kozlenko,
V.S.~Kozlov, A.G.~Krivshich, V.V.~Kulikov,
B.V.~Morozov,  V.M.~Nesterov, D.V.~Novinsky, 
V.V.~Ryltsov, M.E.~Sadler, V.A.~Sakharov, 
D.~Soboyede, A.D.~Sulimov, V.V.~Sumachev, 
D.N.~Svirida,  V.Yu.~Trautman, E.~Walker, S.~Watson. 
\end{center}
}
 \hfill \epsfig{figure=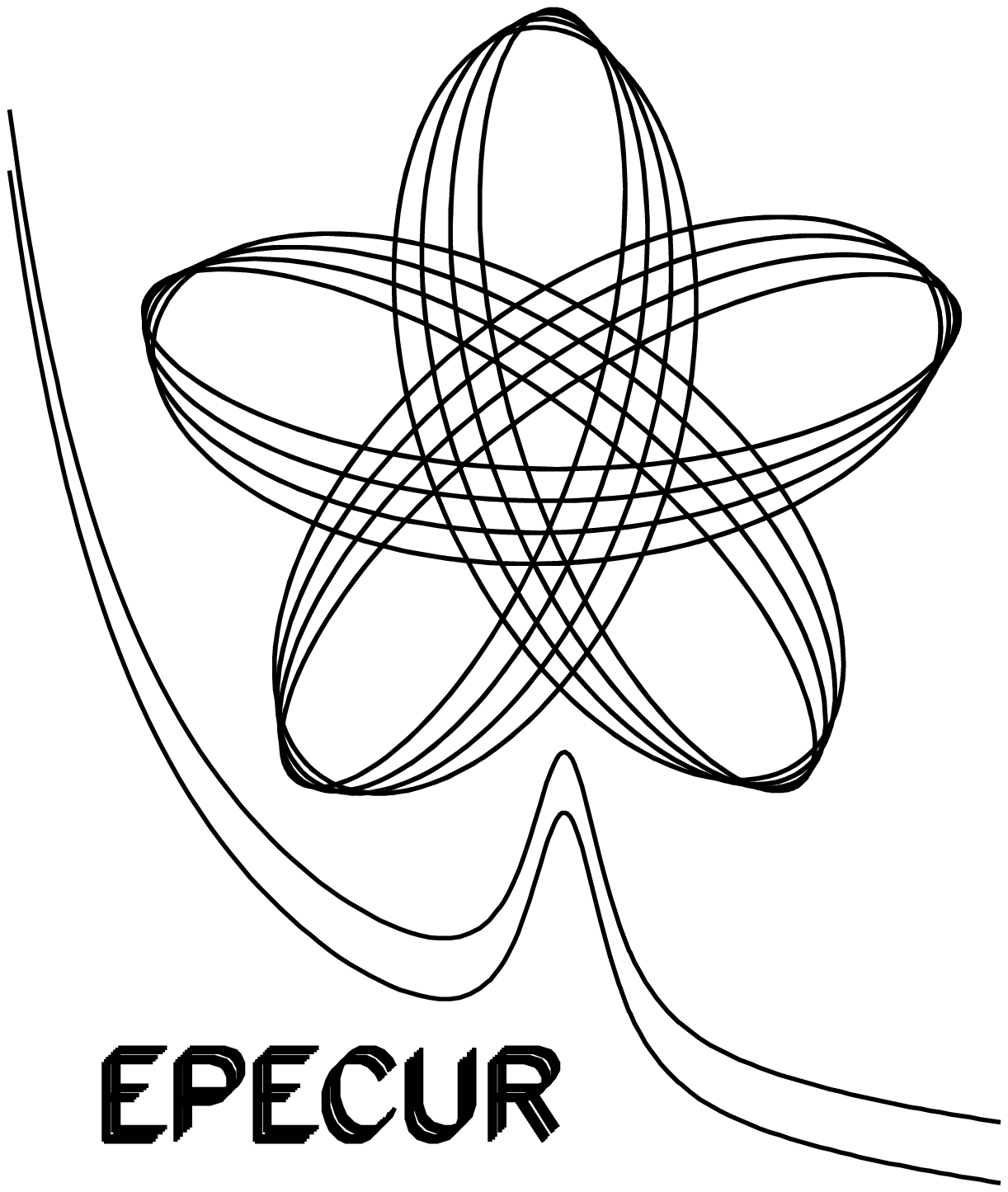, width=0.23\textwidth}
\vskip 35mm
{\bf  \huge Cost Effective Electronics} \\
\vskip 5mm
{\bf  \huge for Proportional and } \\
\vskip 4.0mm
{\bf  \huge Drift Chambers of} \\
\vskip 5mm
{\bf  \huge 'EPECUR' Experiment} \\
\end{center}
\vfill
\centerline{\bf \LARGE \foreignlanguage{russian}{Москва 2006}}

\clearpage
\thispagestyle{empty}
\vspace*{-2em}
\noindent
\foreignlanguage{russian}{УДК 539.1.075 \hfill М-16}

\vskip 0.6em
\noindent
Cost Effective Electronics for Proportional and Drift Chambers of
'EPECUR' Experiment. Preprint ITEP 21-06/

\vskip 0.5em
\noindent
\parbox{\textwidth}{\renewcommand{\baselinestretch}{0.9}%
\normalsize I.G.~Alekseev,  V.A.~Andreev,
P.E.~Budkovsky, E.A.~Filimonov,
V.V.~Golubev, V.P.~Kanavets, M.M.~Kats,
L.I.~Koroleva, A.I.~Kovalev, N.G.~Kozlenko,
V.S.~Kozlov, A.G.~Krivshich, V.V.~Kulikov,
B.V.~Morozov,  V.M.~Nesterov, D.V.~Novinsky, 
V.V.~Ryltsov, M.E.~Sadler, V.A.~Sakharov, 
D.~Soboyede, A.D.~Sulimov, V.V.~Sumachev, 
D.N.~Svirida,  V.Yu.~Trautman, E.~Walker, S.~Watson. \linebreak
M., 2006-23~p. }

\vskip 0.6em
\noindent
\parbox{\textwidth}{\renewcommand{\baselinestretch}{0.9}%
\fontsize{11pt}{13pt}\selectfont
'EPECUR' experiment setup is under construction at the beam line 322
of the ITEP proton synchrotron. The experiment requires several large area
drift chambers to provide reasonable acceptance and fine pitch proportional
chambers for beam particle tracking with total number of electronic channels
of about 7000. New compact and cost effective readout system for these
gaseous detectors was designed, prototyped and tested in the latest two years
based on the modern technologies in analog and digital electronics, as well
as in data transfer protocols. Mass production of the proportional chamber electronics
is close to the end, while the boards for the drift chambers are manufactured
in the amount to equip one 8-plane module. The paper presents the functional
description of the whole DAQ system and its main parts together with some of the
test results as an illustration of the excellent performance of the system.
The appendix contains specific information which may be useful for the system
users or code developers. }

\vskip 0.8em
\noindent
\foreignlanguage{russian}{%
НЕДОРОГАЯ ЭЛЕКТРОНИКА ДЛЯ ПРОПОРЦИОНАЛЬНЫХ И ДРЕЙФОВЫХ 
КАМЕР ЭКСПЕРИМЕНТА "ЭПЕКУР".}

\vskip 0.6em
\noindent
\foreignlanguage{russian}{\parbox{\textwidth}{\renewcommand{\baselinestretch}{0.9}%
\normalsize И.Г.~Алексеев, В.А.~Андреев,
П.Е.~Будковский, В.В.~Голубев,
В.П.~Канавец, М.М.~Кац, Л.И.~Королева,
А.И.~Ковалев, Н.Г.~Козленко, В.С.~Козлов,
А.Г.~Крившич, В.В.~Куликов, Б.В.~Морозов,
В.М.~Нестеров, Д.В.~Новинский, В.В.~Рыльцов,
М.Е.~Садлер, В.А.~Сахаров, Д.~Собоеде,
А.Д.~Сулимов, В.В.~Сумачев, Д.Н.~Свирида,
В.Ю.~Траутман, Е.~Уолкер, С.~Уатсон, Е.А.~Филимонов. \linebreak
М., 2006--23~стр.}}

\vskip 0.5em
\noindent
\foreignlanguage{russian}{\parbox{\textwidth}{\renewcommand{\baselinestretch}{0.9}%
\fontsize{11pt}{13pt}\selectfont
Конструирование экспериментальной установки "ЭПЕКУР"\ идет полным
ходом на 322-м пучке протонного синхротрона ИТЭФ. Для проведения
эксперимента требуется несколько больших дрейфовых камер, чтобы
обеспечить необходимый аксептанс, а также пропорциональные камеры
с мелким шагом проволочек для регистрации частиц пучка. Общее
число необходимых каналов электроники составляет около 7000.
Новая компактная и недорогая система считывания с этих газовых
детекторов была разработана и опробована в течение двух последних
лет. Система использует новейшие достижения технологий в области
аналоговой и цифровой электроники, а также в области протоколов
передачи данных. Массовое производство электроники для пропорциональных
камер близится к завершению, тогда как платы для дрейфовых камер
произведены в количестве, необходимом для оснащения одного
8-плоскостного модуля. В работе представлено функциональное описание
системы сбора данных и ее основных частей, а также результаты различных
тестов как иллюстрация превосходной работы оборудования. В приложении
содержится специфическая информация, которая может быть полезна
пользователю системы или разработчику программного обеспечения.
}} 

\vskip 0.7em
\noindent
{\fontsize{11pt}{13pt}\selectfont
Fig. -- 14, refs. -- 6 names.}

\vfill
\noindent
\hspace{1em}\copyright
\foreignlanguage{russian}
{ Институт теоретической и экспериментальной физики, 2006}

\clearpage
\thispagestyle{empty}
\setcounter{page}{1}
\vspace*{5mm}
\renewcommand{\baselinestretch}{0.85}

\section{Introduction}

\vspace{-0.2em}
'EPECUR' experiment is primarily aimed at the search of the cryptoexotic narrow 
nucleon resonance $N_{\overline{10}}$, the non-strange neutral member
of the pentaquark antidecuplet~\cite{c:EPECUR}. General search for
narrow pion-nucleon resonances is also implied. The method is
the scan of the invariant mass of $\pi^-p$ system in \linebreak {\em s}-channel 
in the region (1610--1770)~MeV with the explicit detection
of $\pi^-p$ and $K^0\Lambda$ final states, which is essentially
a very accurate cross-section measurement of the mentioned processes
in very fine steps in energy.

The setup is under construction at the universal beam line 322
of the \linebreak U-10 synchrotron (ITEP, Moscow). The beam line ideally fits the
experiment requirements due to its very good momentum resolution
of better than 0.1~\%~\cite{c:KATZ} and reasonably high typical intensity 
of $2\cdot10^5$ negative pions per accelerator cycle (every 4 seconds).
The project is a united effort of ITEP (Moscow), PNPI (St. Petersburg)
and ACU (Abilene, USA). The construction started in 2005 from scratch
and the beginning of data taking is scheduled on early 2008.

To reach the goal of the experiment two types of tracking detectors
are required:
\vspace{-0.35em}
\begin{itemize}
\itemsep 0cm
\parskip 0cm
\item[a)] small but fast detectors for incident particle 
tracking and its momentum measurement;
\item[b)] large acceptance high accuracy detectors for the tracking of the
reaction products.
\end{itemize}
\vspace{-0.35em}
Both detectors should have as little matter as possible on the way of the
particles to minimize multiple scattering and maintain the beam momentum
resolution.

Fine pitch proportional chamber is a good solution for a). In addition
such detector provides fast trigger capability, allowing to minimize
the number of scintillation counters in the beam. Two-coordinate
chamber with 1~mm pitch and 200$\times$200~mm$^2$ sensitive area
has 400 sensitive wires.

Wire drift chambers with hexagonal cell structure~\cite{c:SACLAY} 
and 20 mm cell diameter were selected as the best choice for b). Typical drift 
chamber module covers 800$\times$1200~mm$^2$ area and is a stack of 4 submodules
with horizontal, vertical and tilted wire directions. Each submodule
has two planes of drift cells, resulting in approximately 530 sensitive
wires per module.

The setup at its final stage includes 6 proportional chambers, 
5 drift modules of the mentioned size, as well as a twice larger drift
module and 4 smaller modules. This gives 2400 proportional
and approximately 4300 drift sensitive wires and corresponding number
of electronic channels.

Limited funding of the project together with the relatively large
number of required electronic channels put out a challenge to built
a cost effective data acquisition system for the experiment.
To lower the cost the simplest and most compact solution was
implemented using only commercially available components.
Variety of modern integrated circuits, both analog and digital,
together with advanced PCB technology allowed to develop
best quality devices meeting the experiment requirements.
Interconnection between the parts of the DAQ is based on the
wide spread protocols such as USB and Ethernet, allowing application
of standard interfacing modules. 

During 2005-2006 all parts of the DAQ were designed, manufactured
and successfully tested with the prototype detectors. Mass production
of proportional chamber electronics and auxiliary hardware is close 
to the end, while the boards for the drift chambers are in production
in the quantity necessary to equip two prototype submodules.

This paper describes the structure of the 'EPECUR' data acquisition
system, as well as the principles of operation of its main components.
Selected test results are also presented.

\section{'EPECUR' DAQ Structure}

\begin{figure}
\centerline{\epsfig{figure=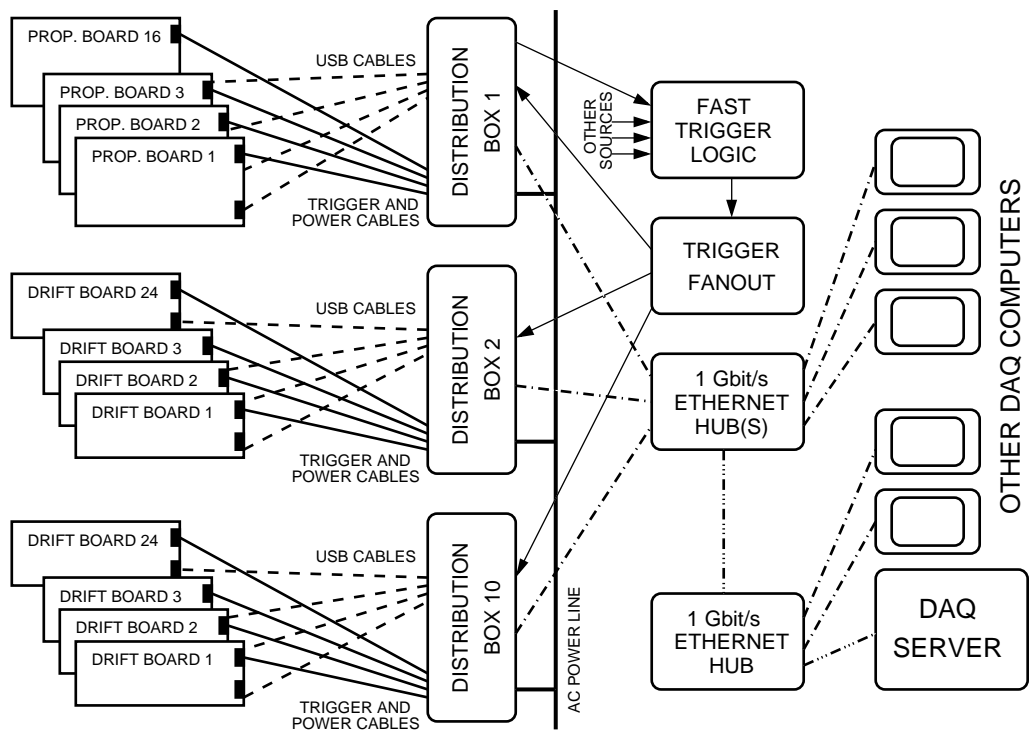, width=0.8\textwidth}}
\caption{'EPECUR' data acquisition system block diagram.}
\label{f:DAQ}
\end{figure}

The data acquisition system is illustrated by Fig.~\ref{f:DAQ} and includes:
\vspace{-0.25em}
\begin{enumerate}
\itemsep 0cm
\parskip 0cm
\item 100-channel proportional boards performing the following functions:
\vspace{-0.4em}
    \begin{itemize}
    \itemsep 0cm
    \parskip 0cm
    \item wire signal amplification and conditioning;
    \item leading edge pulse discrimination;
    \item latching of the signals on the arrival of late trigger by means
of the digital delay line;
    \item encoding of the hit wire numbers into the digital data flow;
    \item transfer of the data through a high speed USB connection;
    \item forming of the fast trigger output as an OR of all wire hit signals;
    \end{itemize}
\vspace{-0.4em}
\item 24-channel drift boards with similar functionality,
but with added time digitization capability and no fast trigger output;
\item locally placed distribution boxes providing power, trigger and
USB connections for the boards, as well as the Ethernet interface 
for the data;
\item DAQ server computer and several workstations for control and
monitoring, connected into a local fast Ethernet network.
\end{enumerate}
\vspace{-0.25em}

Proportional and drift boards are mounted directly on the chamber
frames close to the sensitive wire ends. Each of the boards is connected
to a distribution box with a 4-wire USB cable and a 4-pair UTP 
(Unshielded Twisted Pair)
cable carrying power and differential
trigger signals.

\renewcommand{\baselinestretch}{1.0}

\section{100-Channel Proportional Chamber Board}

\begin{figure}
\centerline{\epsfig{figure=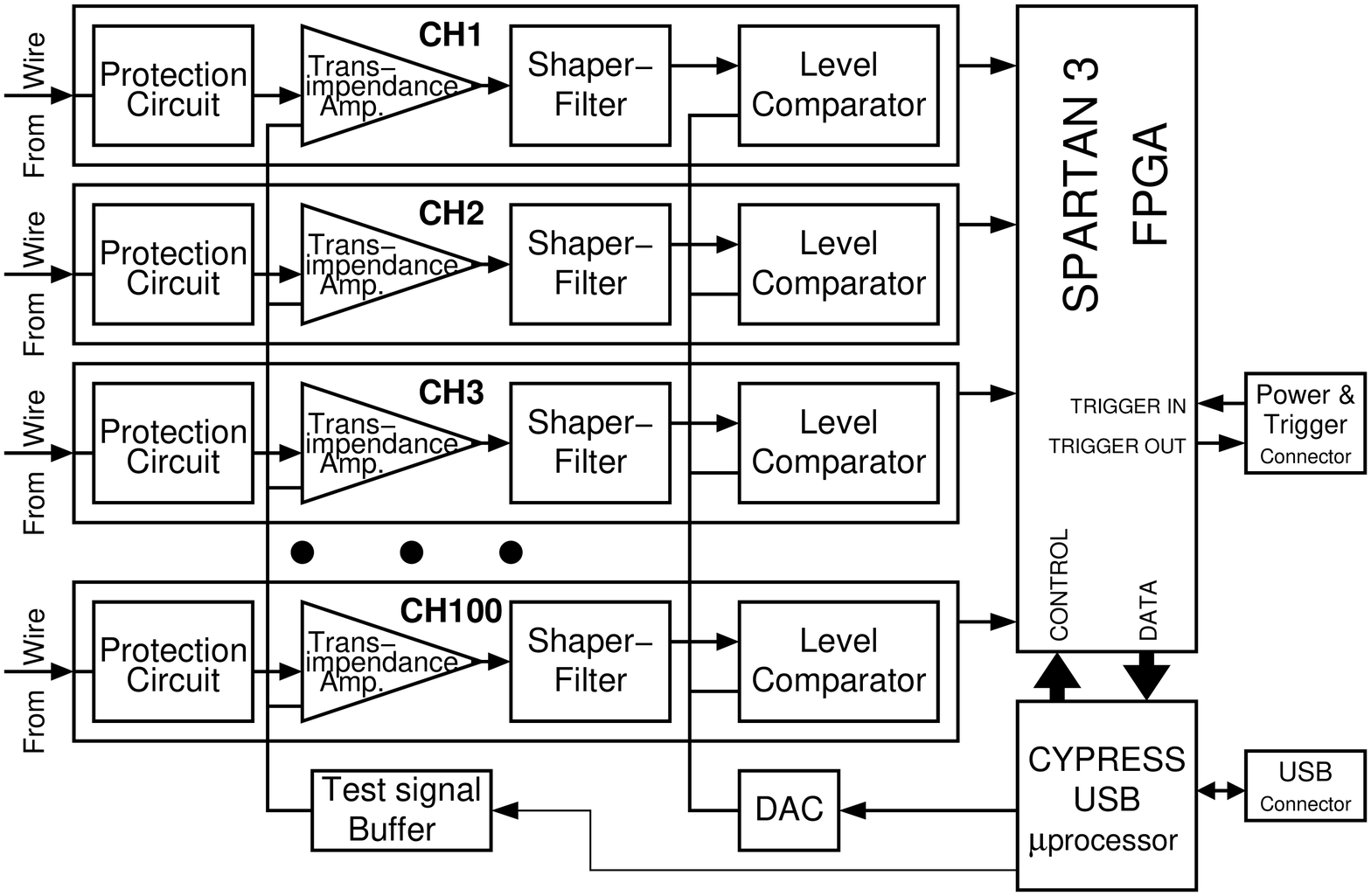, width=0.85\textwidth}}
\caption{Proportional chamber board block diagram.}
\label{f:PROP}
\end{figure}

Proportional chamber board (Fig.~\ref{f:PROP}) 
contains 100 identical
analog channels, Xilinx FPGA chip, Cypress USB microprocessor,
discriminator level DAC and power conditioning components.

The input of each channel is resistively connected to the ground
to provide zero potential to the signal wire. The diode-based
low capacitance protection circuit prevents the input from
over-voltage in case of spark discharge in the chamber. The fast
current sensitive transimpendance amplifier is based on a 1.2~GHz 
IC chip and provides the signal amplification of 5~mV/$\mu$A.
The RC shaper-filter is designed to cut out the long tail of the charge
collection signal and to reduce the high frequency noise component
on the amplifier output.

A fast comparator with TTL output is used to form the leading edge
discriminator. All channels have common discrimination threshold
programmable in  1~mV steps by means of a serial DAC controlled 
by the microprocessor. The amplitude spread of the signals
at the comparator input is in the range of 10~mV to 50~mV. Rather
low amplification allows to avoid oscillation problem and
makes the amplifier stage cost effective. At the same time
a careful choice of the comparator with low offset voltage
of several mV is necessary.

\begin{figure}[!b]
\centerline{\epsfig{figure=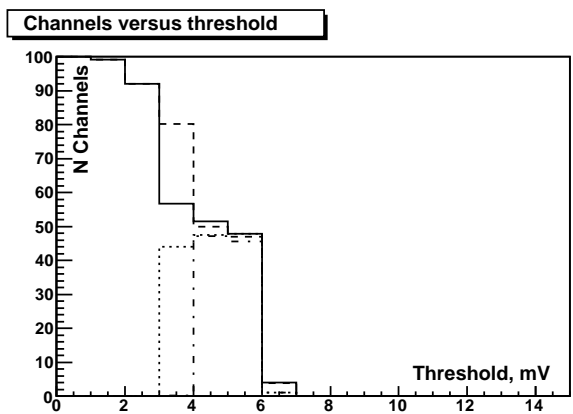, width=0.5\textwidth}}
\caption{Typical results of the proportional board threshold spread test.}
\label{f:THRESH}
\end{figure}

Figure~\ref{f:THRESH} shows the typical results of a board threshold spread
test. Due to the bias currents flowing into the comparator inputs, the
actual discriminator level is always several mV below the common voltage, 
applied to the negative comparator inputs. The software can read each 
channel level state at various DAC settings and thus count the number 
of trigged channels (with no input signal) as a function 
of applied threshold voltage. Solid and dashed lines present the number
of trigged channels when the threshold voltage is changed in steps of 1~mV
from 0 to 15~mV and back, correspondingly. The range of 3 to 5~mV is the 
region of instability, where many of the channels are very close to their
thresholds and start switching synchronously causing positive feedback from
the TTL output to the amplifier input. This is the reason why the
solid and dashed lines do not coincide. The software can also identify
the region where the oscillating switching of the comparator outputs occur;
the half number of oscillating channels is presented by the dash-dotted and dotted
lines for ascending and descending threshold voltages.
As it's seen from Fig.~\ref{f:THRESH}, 
at threshold voltages above 7~mV all 100 channels are not trigged and ready
to accept small signals, while the total threshold spread is 6~mV.

\begin{figure}
\centerline{\epsfig{figure=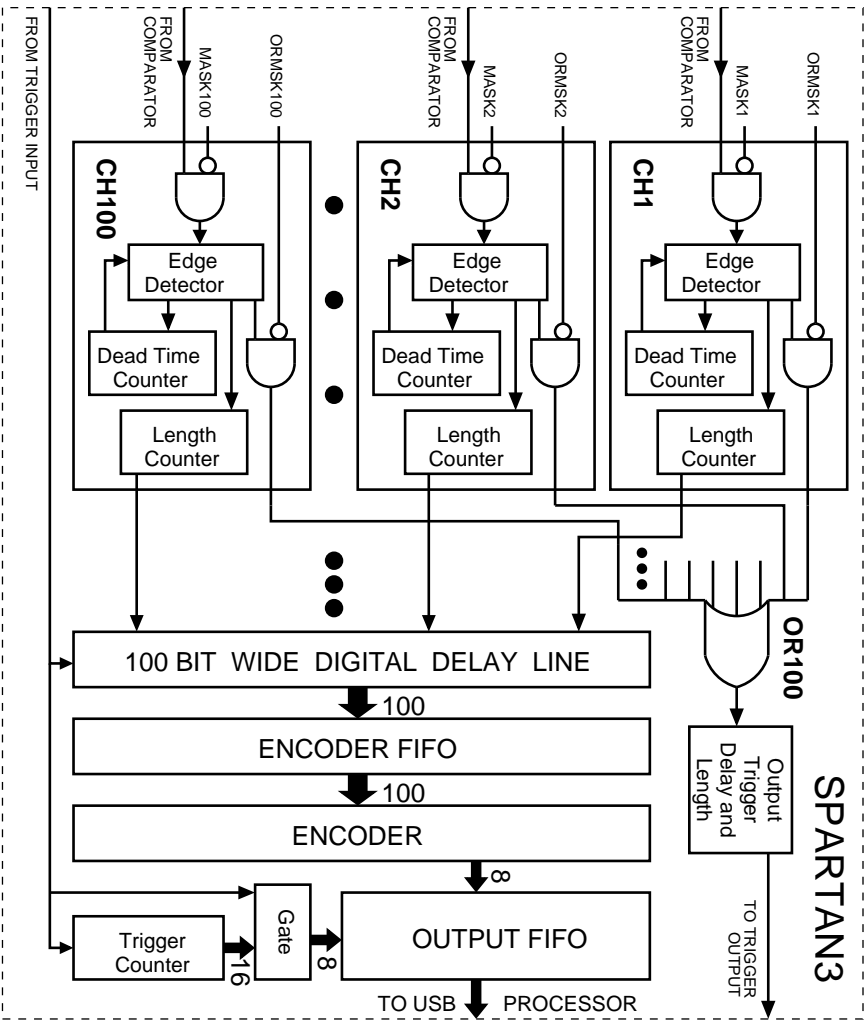, width=0.8\textwidth}}
\caption{Proportional board FPGA block diagram.}
\label{f:PRFPGA}
\end{figure}

All comparator outputs are tied to the inputs of the 
Spartan-3~\cite{c:XILINX} Field Programmable
Gate Array (FPGA) chip and all subsequent signal processing is done
digitally inside this chip. Figure~\ref{f:PRFPGA} illustrates the
the idea of the configuration implemented in th FPGA.

 The leading edge of the signal 
from the comparator is captured by the 'edge detector' and
starts two counters. The 'length counter' is intended for
producing a signal of several clock 
periods\footnote{The internal clock frequency of the FPGA is 96~MHz,
corresponding clock period is about 10~ns.}
which is then put to
the digital delay line. The 'dead time counter' blocks
the further acceptance of signals by the channel for a certain amount of time.
This is done to suppress 'ringing' on the input line caused
by the crossing of the comparator threshold by the long and
slowly changing charge collection tail. Both counters are separately
programmable, but to the same values for all channels (see Table~\ref{t:PRXREG}).
A pulse of one clock period from each channel is used to form the
grand OR trigger output. The duration and the delay of the
signal on this output are also programmable in one clock period steps.
A channel may be completely disabled, or its participation in the 
grand OR may be forbidden, by setting corresponding MASK or ORMSK bits.

The state of all 100 'length counter' outputs  is continuously piped through 
the digital delay line. On the (late) arrival of the actual system trigger
a time slice is retrieved from a certain (programmable) depth of the delay line,
representing the state of all channels on the time of the event
which caused the trigger. This 100-bit wide time slice is stored
to the encoder FIFO which provides buffering and ensures dead-timeless
operation. The encoder converts the nonzero bit positions in the slice into the
sequence of their numbers and directs it to the output FIFO preceded
by some control information including the sequential trigger number
(see Table~\ref{t:PRFMT} for details).

The data is then transferred to the USB microprocessor~\cite{c:CYPRESS} and buffered
there to form an USB packet~\cite{c:USB} which can be retrieved on the request
from the host computer. Another data line is used by the processor
to load and read the programmable registers inside the FPGA. The 
microprocessor is also responsible for generating the test pulse
which is distributed to each analog channel input through a small
capacitance.

The microprocessor code and the FPGA configuration
firmware is not stored on the board thus eliminating additional
cost expenses for the large flash memory chip and saving the board space.
Instead, on power up the bootloader hardcoded into the
microprocessor retrieves the application code through the USB from
the host computer. In turn, this code provides the possibility
to configure the FPGA chip with the firmware downloaded
by the USB host. In addition such approach gives an extreme
flexibility of changing the microprocessor code as well as the
FPGA configuration. A very small I$^2$C flash memory is used to keep
the board identification and serial numbers which are programmed 
at the production time. 

The summary of USB requests served by the application code is
presented in Table~\ref{t:PRUSB}.

\setlength{\unitlength}{\textheight}
\begin{figure}
\begin{picture}(0.6, 0.92)(0,-0.02)
\put(0.3,0.036){\makebox(0.1,0.1)[lb]{\epsfig{figure=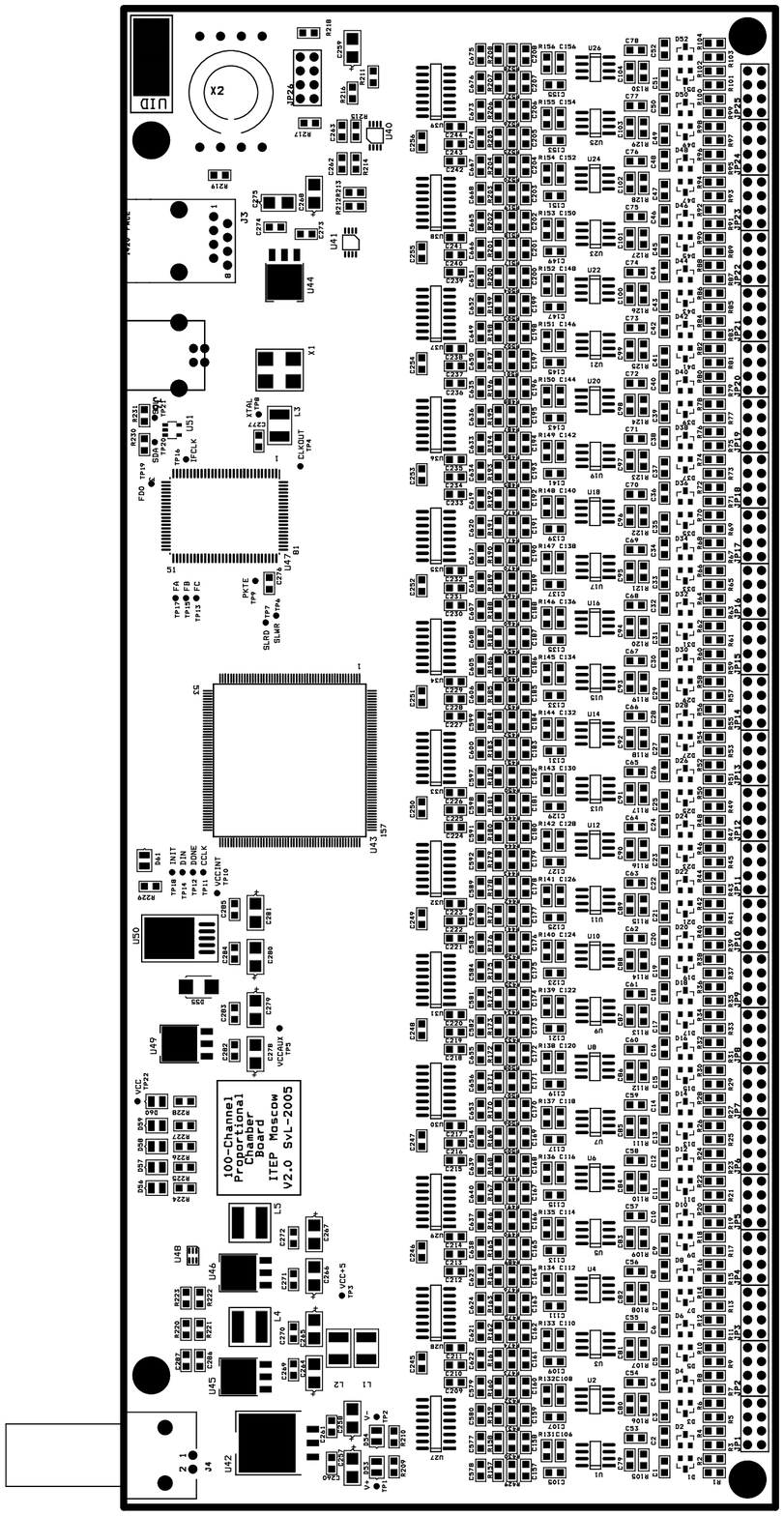, height=0.78\textheight}}}
\put(0,0){\makebox(0.1,0.1)[lb]{\epsfig{figure=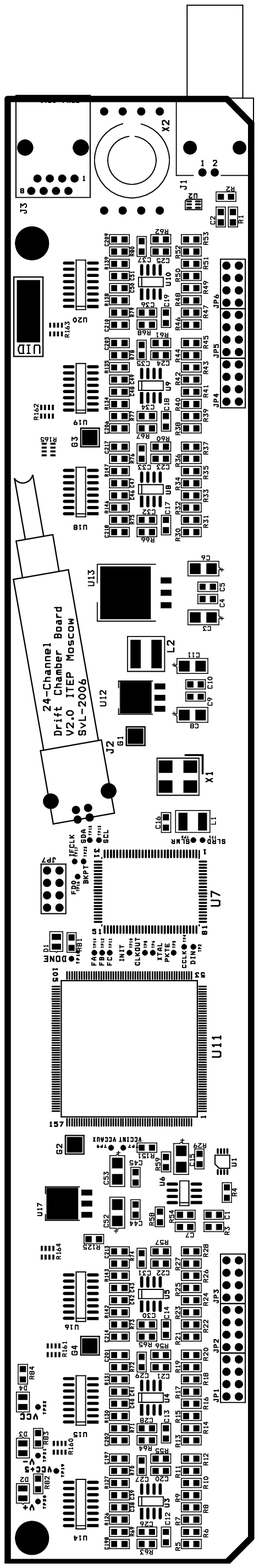, height=0.9\textheight}}}
\put(0.18, 0.35){\parbox{0.2\textwidth}{\Large \sf \centerline{SPARTAN3} \centerline{FPGA}}}
\put(0.2, 0.35){\vector(-2,-1){0.12}}
\put(0.326, 0.38){\vector(4,1){0.12}}
\put(0.18, 0.44){\parbox{0.2\textwidth}{\large \sf \centerline{CYPRESS USB} \centerline{$\mu$PROCESSOR}}}
\put(0.18, 0.44){\vector(-2,-1){0.09}}
\put(0.33, 0.47){\vector(1,1){0.08}}
\put(0.18, 0.53){\parbox{0.2\textwidth}{\large \sf \centerline{USB} \centerline{CONNECTOR}}}
\put(0.175, 0.515){\vector(-2,-1){0.13}}
\put(0.28, 0.56){\vector(4,3){0.095}}
\put(0.20, 0.85){\parbox{0.2\textwidth}{\large \sf \centerline{POWER\&TRIGGER} \centerline{CONNECTOR}}}
\put(0.165, 0.87){\vector(-3,-1){0.13}}
\put(0.28, 0.825){\vector(3,-4){0.10}}
\put(0.21, 0.645){\parbox{0.2\textwidth}{\large \sf \centerline{ANALOG} \centerline{CHANNELS}}}
\put(0.22, 0.65){\vector(-4,-1){0.13}}
\put(0.22, 0.65){\vector(-4,1){0.13}}
\put(0.22, 0.65){\vector(-4,3){0.13}}
\put(0.34, 0.65){\vector(4,1){0.27}}
\put(0.34, 0.65){\vector(4,-1){0.27}}
\put(0.155, 0.72){\parbox{0.2\textwidth}{\small \sf CHAMBER \\ CONNECTOR}}
\put(0.155, 0.14){\parbox{0.2\textwidth}{\small \sf CHAMBER \\ CONNECTOR}}
\put(0.4, 0.0){\parbox{0.3\textwidth}{\small \sf CHAMBER CONNECTOR}}
\put(0.6, 0.01){\vector(1,1){0.09}}
\setlength{\unitlength}{0.676cm}
\put(14.5, 29.5){\line(1,0){10.0}}
\put(14.5, 29.7){\line(1,0){10.0}}
\put(24.5, 29.5){\line(0,1){0.4}}
\put(19.5, 29.5){\line(0,1){0.4}}
\put(14.5, 29.5){\line(0,1){0.4}}
\put(14, 30){0 cm}
\put(19, 30){5 cm}
\put(23, 30){10 cm}
\linethickness{0.135cm}
\multiput(14.5, 29.6)(2,0){5}{\line(1,0){1}}
\end{picture}
\caption{Proportional (right) and drift (left) chamber boards. Chamber
connectors are not mounted on both board views.}
\label{f:BOARDS}
\end{figure}

The board dimensions are 275$\times$120~mm$^2$ (Fig.~\ref{f:BOARDS}) 
and the connection to the chamber is done via the 200~pin two row 
90$^o$ male header connector (all pins in the second row are the signal ground). 
Several LED indicators (see Table~\ref{t:PRLED}) visualize the board status.
Power requirements are listed in the
Table~\ref{t:PRPWR}. The power is supplied to the board through
an RJ45 8~pin connector (Table~\ref{t:PRCONN}) which also provides 
the trigger connection.
Natural convection airflow is enough at room temperature for the board
cooling in any position, but forced ventilation is required if the board
is placed in a closed volume.

\begin{table}[h]
\caption{Power requirements for the 100-channel proportional chamber board.}
\label{t:PRPWR}
\begin{center}
\begin{tabular}{|r|r|r|c|}
\hline
Supply & Range & Current & Power dissipation \\
\hline
+6~V & 5.5--6.5~V & 1.4~A & 8.4~W \\
\hline
--6~V & --(5.5--6.5)~V & 1.3~A & 7.8~W \\  
\hline
+4~V & 3.6--4.3~V & 0.4~A & 1.6~W \\
\hline
\multicolumn{3}{|r|}{Total power dissipation:} & 17.8~W \\
\hline
\end{tabular}
\end{center}
\end{table}

\vspace{-2em}
\section{24-Channel Drift Chamber Board}

The structure of the drift chamber board is very similar to that
of the proportional one, discussed in the previous section (Fig.~\ref{f:PROP}). 
The main differences are:
\begin{itemize}
\itemsep 0cm
\parskip 0cm
\item only 24 analog channels per board;
\item faster transimpendance amplifier;
\item completely different FPGA configuration with individual channel
timing capability;
\item additional circuity for temperature and power monitoring;
\item smaller board size.
\end{itemize}

\begin{figure}
\centerline{\epsfig{figure=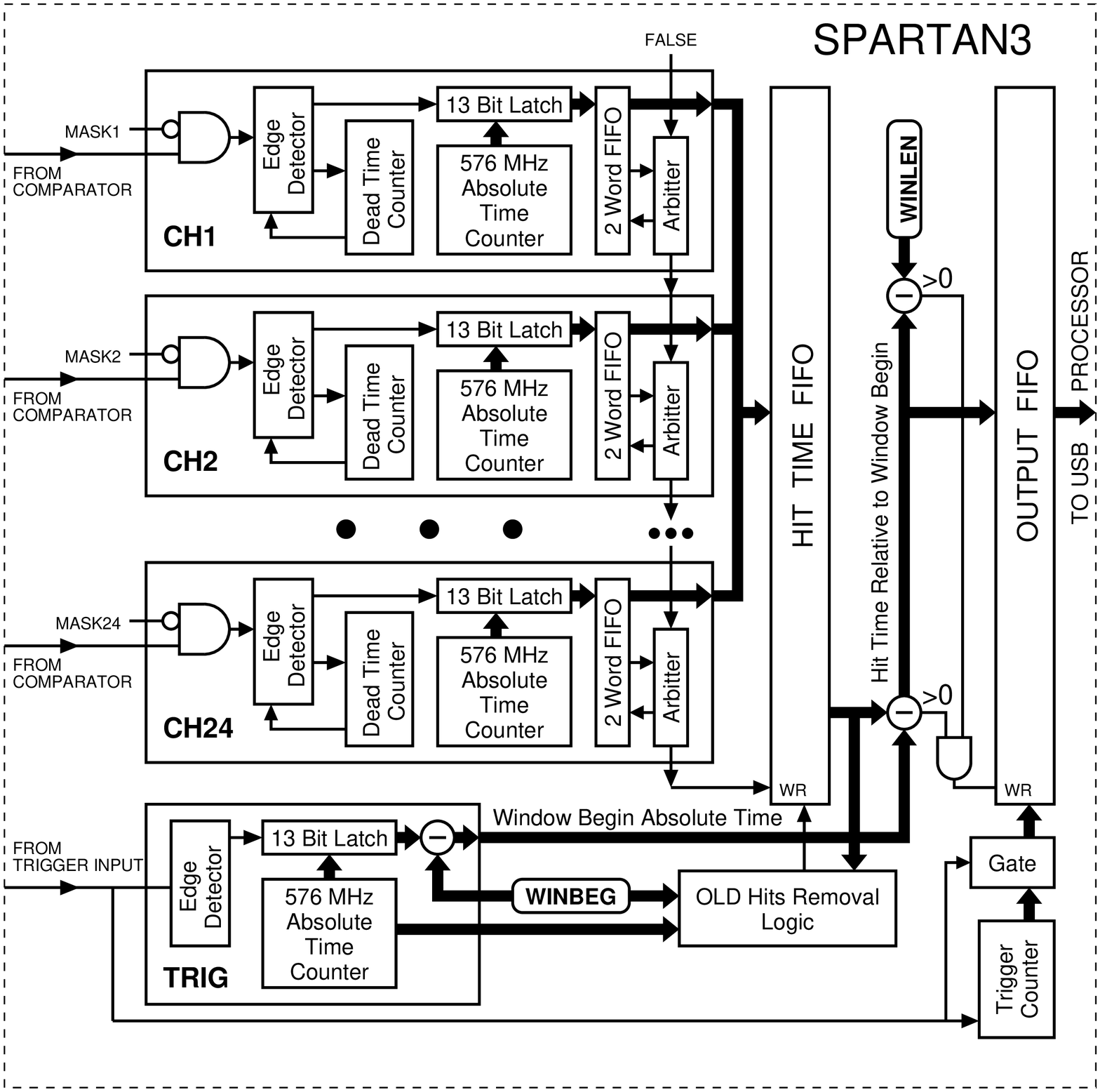, width=0.85\textwidth}}
\caption{Drift board FPGA block diagram.}
\label{f:DRFPGA}
\end{figure}

The FPGA configuration for the drift chamber board is presented
in Fig~\ref{f:DRFPGA}. The input stage of each channel is similar to that
of the proportional design and includes 'edge detector' and 'dead time counter'.
On the detection of the signal the value of the 13-bit absolute time counter
is latched and stored to an intermediate '2 word FIFO'. 
The equivalent frequency of the counter
is 576~MHz, corresponding to 1.74~ns time measurement unit. Each channel
has its locally placed absolute time counter to minimize the connection
delays on the FPGA crystal. All counters are synchronized on the system
initialization and at the beginning of each accelerator cycle period. Nonempty
intermediate FIFOs store their content to the main 'hit time FIFO'
arbitrating on the 'channel number priority' basis. Channel number
is stored to the main FIFO together with the absolute time value latched
by this channel.

\begin{figure}
\centerline{\epsfig{figure=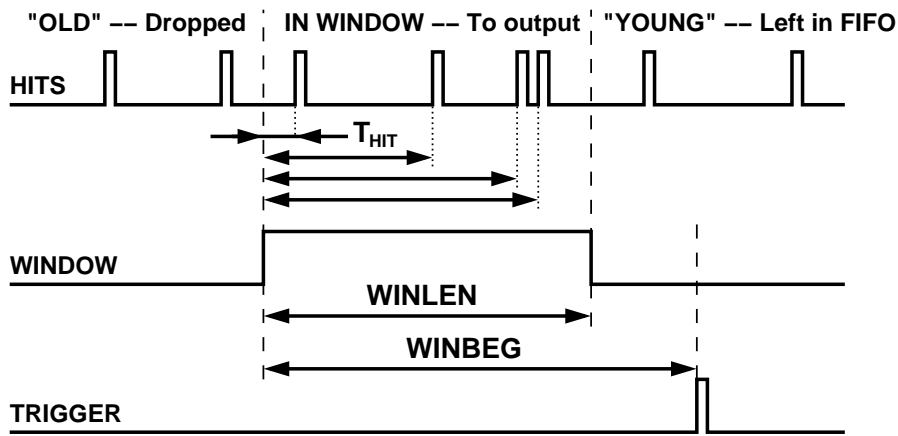, width=0.7\textwidth}}
\caption{Drift board timing.}
\label{f:DRTIME}
\end{figure}

Only those channel hits that fall into a certain window are
processed. Two programmable registers (Table~\ref{t:DRXREG})  
are responsible for the definition of this window: {\tt WINBEG} 
designates the beginning of the window prior to the arrival of
the system trigger, while {\tt WINLEN} is the window length 
(see Fig~\ref{f:DRTIME}).

The absolute time value of the oldest (first in the queue) hit
is always present on the main FIFO output. The running value of the
'age threshold' is continuously calculated by the 'old hits removal
logic' as the absolute time decreased by the {\tt WINBEG} setting 
(actually this is 'window begin' given the trigger arrives 
at the current moment). The two values are continuously 
compared and the hit is removed from the FIFO as soon as 
it becomes older than the 'age threshold'. 
Thus the oldest hit in the FIFO on the moment of actual
trigger arrival will be the first hit falling into the 
trigger window. The procedure also ensures that the FIFO
is never overflown in case no triggers follow the wire signals. 

On the detection of the system trigger processing of the channel 
signals arrived by this time starts. The trigger arrival time
is digitized by the separate channel also in terms of the absolute time.
The beginning of the actual window is calculated and this value
is then subtracted from the consequent hit values from the FIFO
to form the hit time values $T_{HIT}$, relative to the window begin.
These resulting values are passed to the output FIFO preceded
by the sequential trigger number.
The procedure stops and the trigger is considered to be served
when the main FIFO is empty or a hit value later than the window end
is encountered. The hits left in the FIFO will either
become 'old' and be removed or picked up by the next trigger window.

Processing of each hit in the window takes one CLK2 
period\footnote{CLK2 frequency is 48~MHz, corresponding period is about 20~ns}, 
so the trigger service time is proportional to the number of
hits in the window. During this service time other triggers appearing
on the input do not start the hit selection procedure, but increment
the trigger counter. During the data taking, if even some triggers
are lost, probably differently in different boards, the trigger 
number preceding the data of a certain event will always be
the same in all boards. 

In case of frequent triggers, the trigger windows of the consequent triggers
may overlap. In this case the earlier trigger will sweep the hits from 
the main FIFO, which might also correspond to the later one. This possibility
should be taken into account in the data processing.

The further data transferring in the drift chamber board is 
essentially the same as in the proportional one, 
except that the structure of the data flow is different reflecting
the timing capability (Table~\ref{t:DRFMT}).

The board dimensions are 300$\times$50~mm$^2$ (Fig.~\ref{f:BOARDS}), connection
to the chamber is done via the two 24~pin two row 90$^o$ male header connectors
(as in the proportional design, every second pin is signal ground). 
LED indicators are similar to those of the proportional board 
(Table~\ref{t:PRLED}). Power requirements are listed in the
Table~\ref{t:DRPWR}. Power and trigger connection
is the same as in the proportional design (Table~\ref{t:PRCONN}), 
so that the same cables
can be used for both types of the boards. Power dissipation
is more than 3 times less than that of the proportional board,
yet forced airflow should be used for cooling if many
of the boards are placed in a closed volume.

\begin{table}[h]
\caption{Power requirements for the 24-channel drift chamber board.}
\label{t:DRPWR}
\begin{center}
\begin{tabular}{|r|r|r|c|}
\hline
Supply & Range & Current & Power dissipation \\
\hline
+6~V & 5.5--6.5~V & 0.33~A & 2.0~W \\
\hline
--6~V & --(5.5--6.5)~V & 0.31~A & 1.9~W \\  
\hline
+4~V & 3.6--4.3~V & 0.4~A & 1.6~W \\
\hline
\multicolumn{3}{|r|}{Total power dissipation:} & 5.5~W \\
\hline
\end{tabular}
\end{center}
\end{table}

\section{Distribution Box}

In the framework of 'EPECUR' DAQ system the data is transferred 
from the chamber boards by means of universal serial bus (USB)~\cite{c:USB}.
This approach has many obvious advantages, based on the fact
that USB is well developed and widely spread:
\begin{itemize}
\itemsep 0cm
\parskip 0cm
\item high transfer speed 480~Mbit/s;
\item thin 4 wire connection cable;
\item up to 127 'devices' on one branch;
\item 'device' side controller chips are readily available and
relatively cheap;
\item software development tools are well developed for both
'host' and 'device' sides;
\item many standard items are commercially available (i.e. cables,
connectors, hubs etc.).
\end{itemize}

The main disadvantage of USB is the 5~m limit on a single connection
cable length, while the experiment requires transferring of the data
for distances up to 100~m.

To solve this problem the distribution boxes were developed, which
are placed locally near the chambers and serve as an intellectual interface
collecting data from the boards through the USB and sending them to the server
computer by means of fast Ethernet.

At the same time power and trigger signals need to be delivered
to the boards. Placing of the power supplies and trigger distribution
logic into the same box with the data interface allows to make the 
system most compact.

\begin{figure}
\centerline{\epsfig{figure=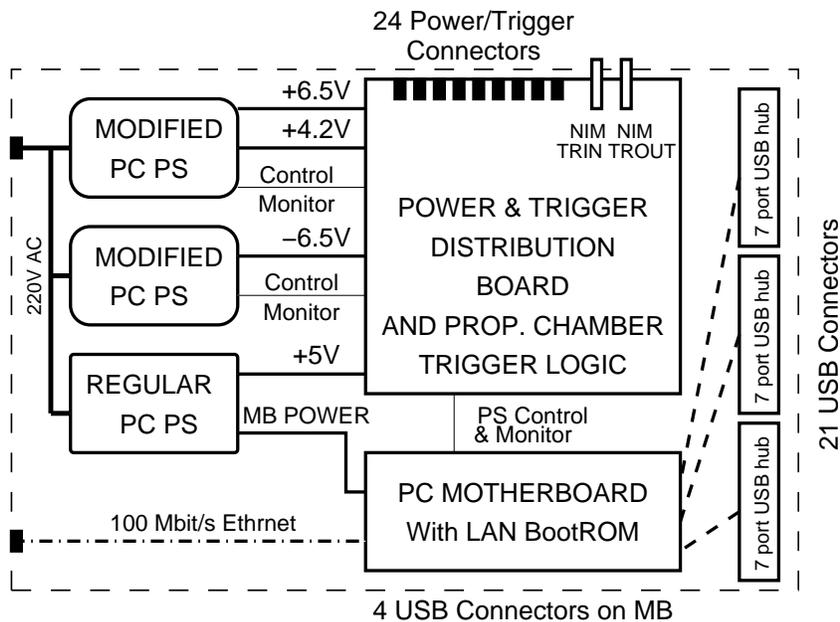, width=0.65\textwidth}}
\caption{Distribution box block diagram.}
\label{f:DISTR}
\end{figure}

To achieve minimum cost most of the components used in the distribution
box are just commercially available blocks. The distribution box contents
is illustrated in Fig.~\ref{f:DISTR} and includes:
\begin{itemize}
\itemsep 0cm
\parskip 0cm
\item standard Pentium based motherboard with USB, Ethernet LAN, parallel
and serial controllers on it and enabled 'LAN boot' feature;
\item regular PC power supply (PS) providing power to the motherboard
and control circuits on the distribution board;
\item two equally modified standard 300~W PC power supplies, capable
of up to 25~A current on main low voltage supply lines;
\item 3~USB hubs, 7~port each, allowing up to 25~USB connections together
with 4~USB ports on the motherboard;
\item specially designed distribution board, performing
the following functions:
    \begin{itemize}
    \itemsep 0cm
    \parskip 0cm
    \item power distribution for up to 24 boards through RJ45 connectors;
    \item additional power filtering and reverse voltage protection;
    \item conversion of the NIM system trigger signal to LVDS and its 
    multiplication for individual boards;
    \item reception of the LVDS grand OR triggers from the proportional
    chambers, performing majority coincidence function on them and
    conversion of the resulting signal to NIM for further use by the
    fast system trigger logic;
    \item conditioning of the PS status signals and making them
    available for the motherboard through the serial port modem lines;
    \item providing remote board power control (on/off) using another
    serial port line;
    \item programming of the fast Xilinx CPLD~\cite{c:CPLD}, which performs majority
    coincidence, by means of the motherboard parallel port;
    \item visible LED indication of all power lines and trigger signals.
    \end{itemize}
\end{itemize}
\vspace{-0.25em}

The distribution box is capable of serving up to 24 drift chamber
boards, but only up to 16 proportional chamber boards due to the
current limit of the power supplies. The box is connected to the 
outer world by means of only 4~cables:
\vspace{-0.25em}
\begin{itemize}
\itemsep 0cm
\parskip 0cm
\item AC line cable;
\item fast Ethernet cable;
\item two coaxial cables, transferring NIM triggers to and from the box.
\end{itemize}
\vspace{-0.25em}

The box dimensions are $500\times 200\times350$~mm$^3$ (W$\times$H$\times$D).
The power consumption from the AC line strongly depends on the number
of boards connected to the box but does not exceed 400~W. Three power supply
fans provide sufficient cooling airflow inside the whole box.

\section{Selected Test Results}

Both types of boards and the whole DAQ system underwent a lot of
various tests both on laboratory tabletop and in real beam conditions.
The detailed description of the test procedures is beyond the scope
of this paper, yet some of the test results would give a good
illustration of the system performance.

The proportional boards were tested with 1~mm pitch 200$\times$200~mm$^2$
proportional chambers both with a radioactive $\beta$-source and with
proton beams of the ITEP synchrotron and of the PNPI cyclotron. 
Two gas mixtures were tried: Ar(67\%)CO$_2$(33\%) and
Ar(75\%)C$_4$H$_{10}$(24.7\%)CF$_3$Br(0.3\%),
so called 'magic mixture'. Both variants showed excellent performance
in various beam conditions, except that the first gas mixture
is efficient at potential plane voltages about 300~V higher than
that for the 'magic mixture'.

The figures below illustrate the typical performance of the
proportional system under the following conditions:
\vspace{-0.25em}
\begin{itemize}
\itemsep 0cm
\parskip 0cm
\item collimated $\beta$-source;
\item chamber load $1\cdot10^5$ $e$/s;
\item trigger rate $2.5\cdot10^3$ s$^{-1}$;
\item 'magic' gas.
\end{itemize}
\vspace{-0.25em}

\begin{figure}
\centerline{\epsfig{figure=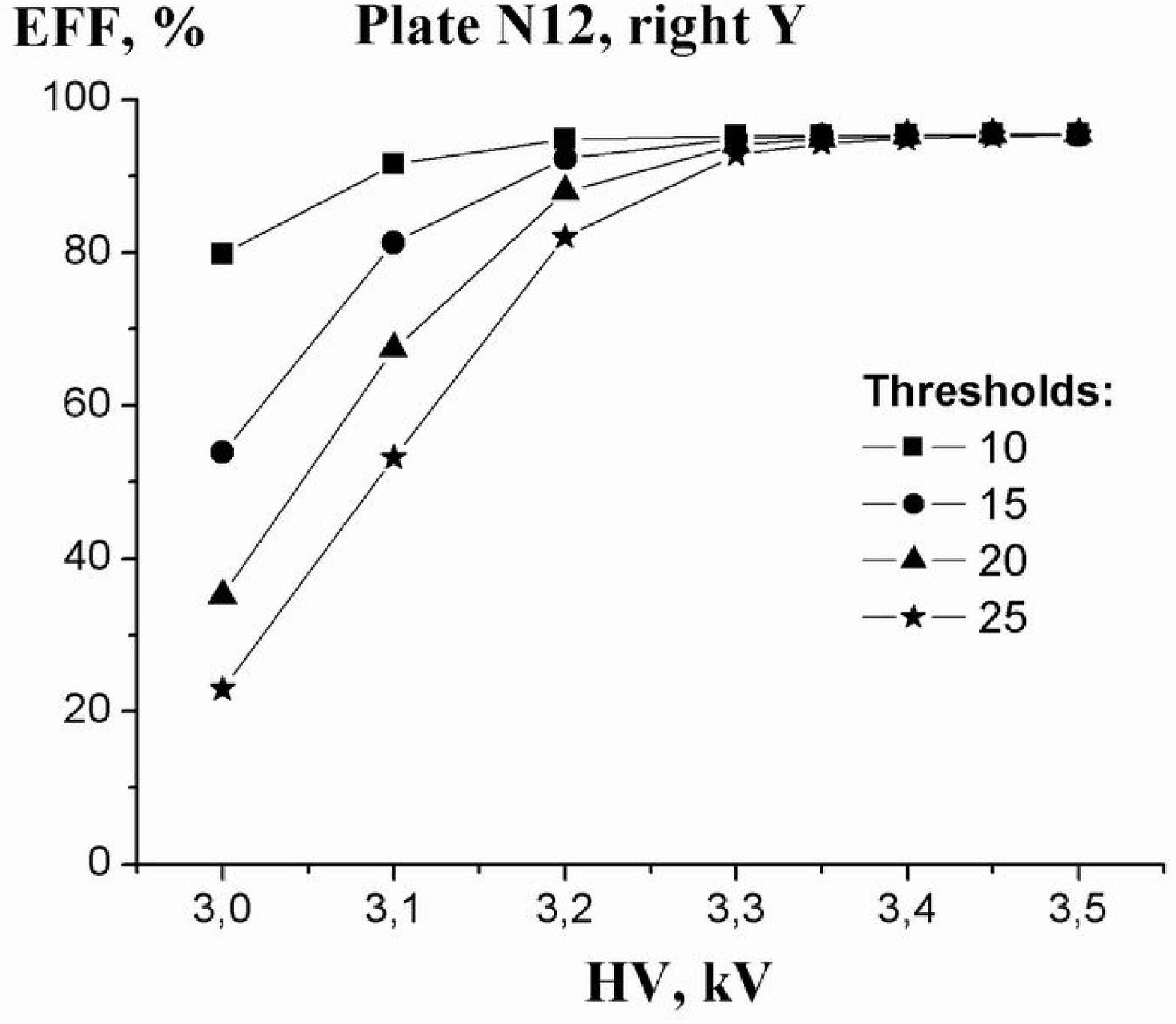, width=0.6\textwidth,
height=0.32\textheight}}
\caption{Proportional chamber efficiency as a function of high voltage
and discriminator threshold setting.}
\label{f:PREFF}
\end{figure}

Figure~\ref{f:PREFF} shows the efficiency of the chamber as a function of the
applied high voltage at various discriminator threshold settings. The
maximum value is lower than 100\% due to the relative geometry of the
chamber and the trigger scintillation counter. Yet a reliable plato 
of 100--200~V can be observed for all threshold values.

\begin{figure}[!b]
\centerline{
\begin{tabular}{cc}
\raisebox{0.62mm}{
\epsfig{figure=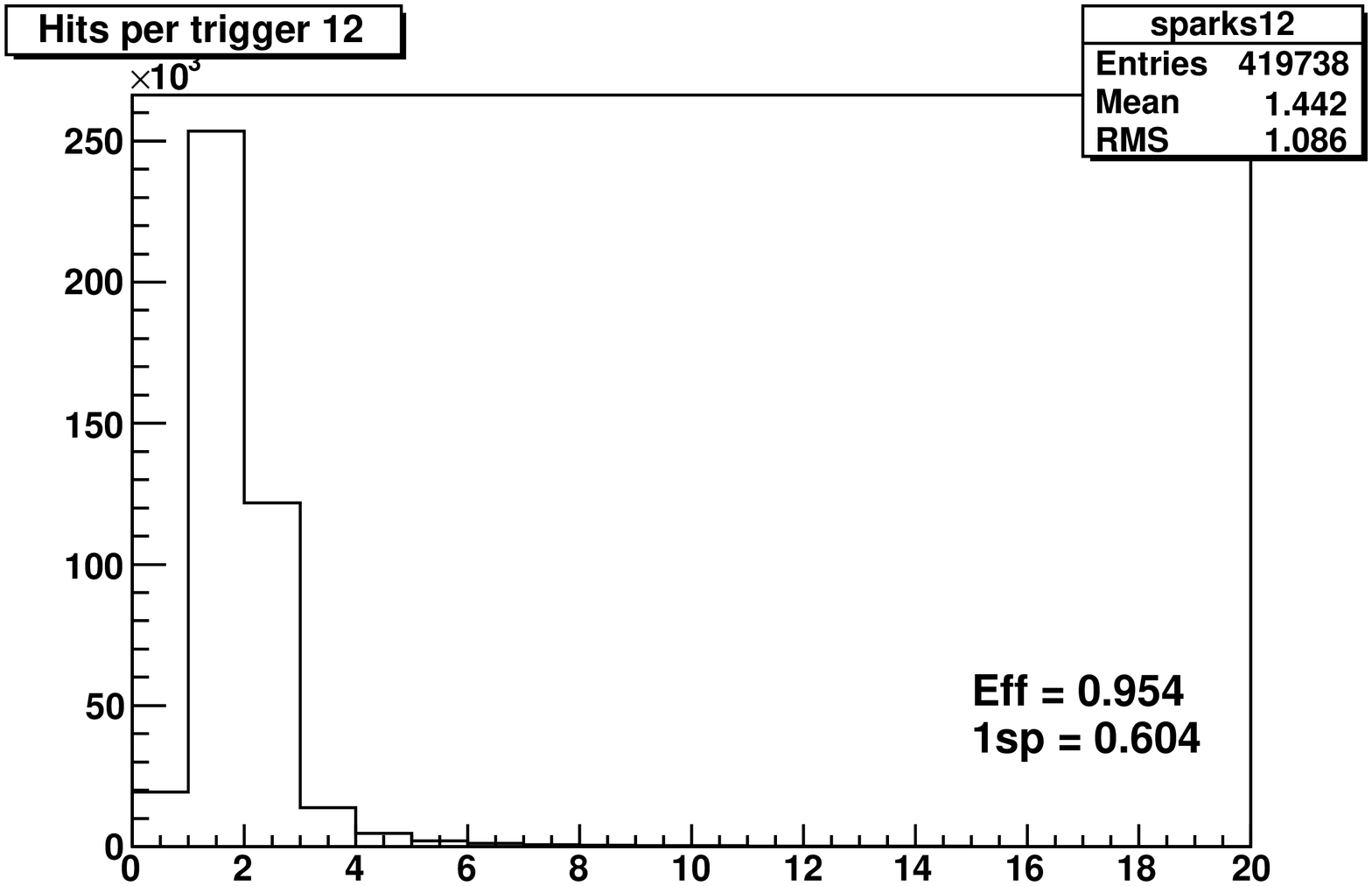, width=0.45\textwidth,
height=0.217\textheight}}&
\epsfig{figure=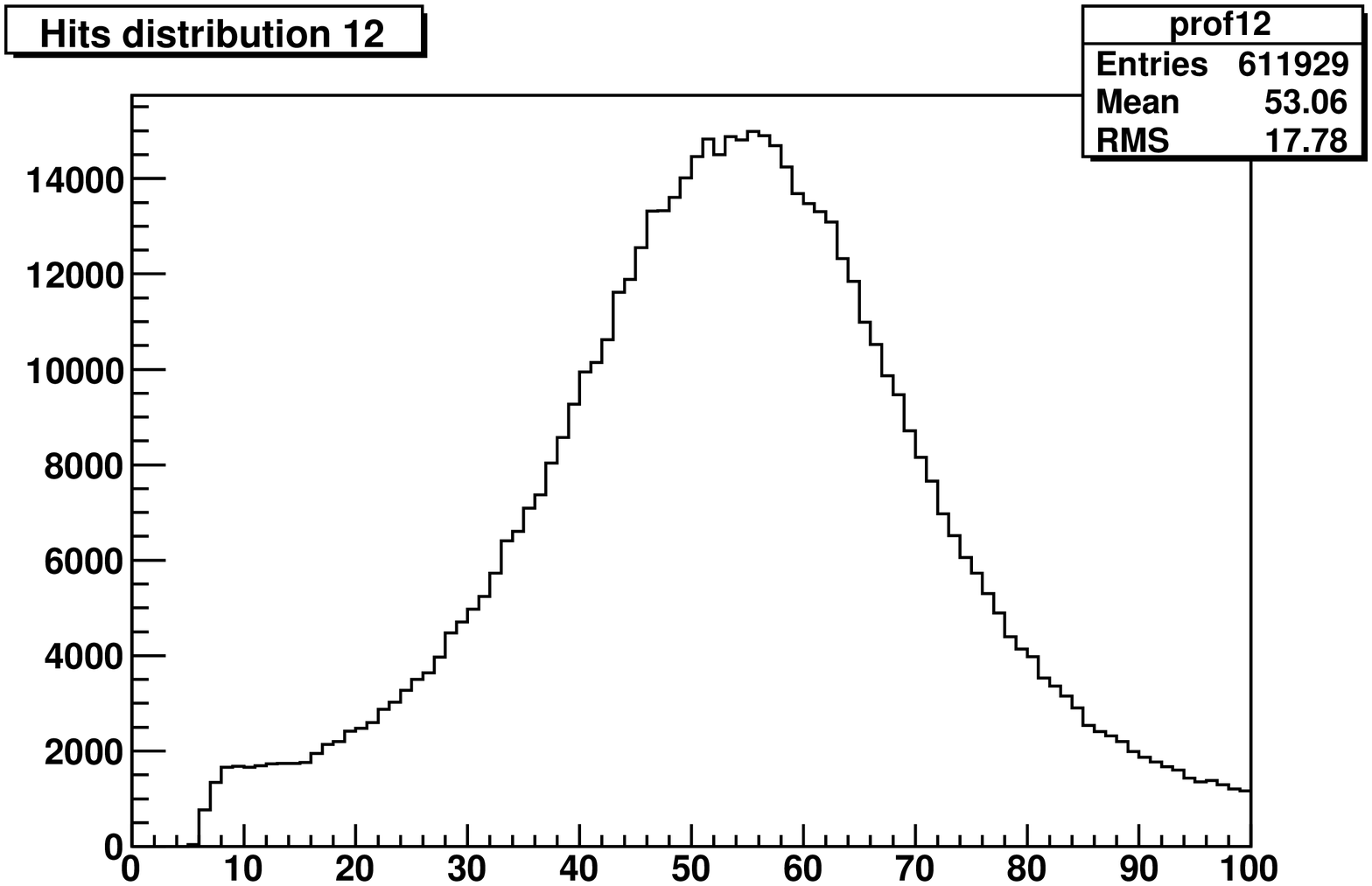, width=0.5\textwidth,
height=0.22\textheight} \\
a) & b) \\
\end{tabular}
}
\caption{Number of hits per trigger in the proportional chamber (a) and
$\beta$-source profile distribution (b); 
HV=3.5~kV, discriminator threshold is 25~mV.}
\label{f:PRNHPT}
\end{figure}

The distribution of the number of hits detected in the chamber for
a given trigger is shown in Fig.~\ref{f:PRNHPT}a. Some events in zero bin
account for the geometrical inefficiency. Nearly 60\% of particles
produce exactly one hit in the chamber. About 30\% of the events
give two hits, which is natural if charge division between adjacent
wires for inclined tracks is taken into account. Figure~\ref{f:PRNHPT}b is
obtained at the same conditions and represents the profile distribution
of the source beam. Only one 100-channel board is active and the source is
centered in the half-plane, corresponding to this board. The leftmost
6 wires are insensitive by the chamber design.

1~mm pitch proportional chambers are considered to be difficult to work with,
compared to commonly used 2~mm ones. Yet the designed system shows
excellent performance even in this case.

\begin{figure}
\centerline{\epsfig{figure=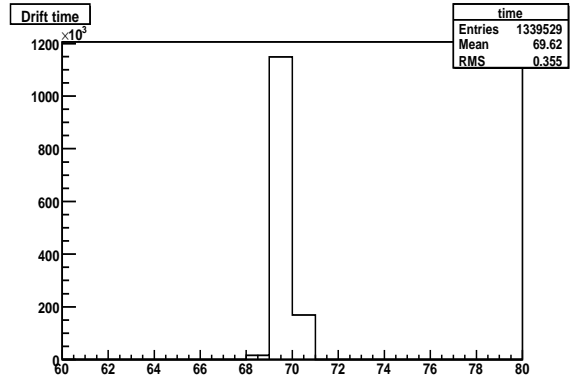, width=0.45\textwidth,
height=0.22\textheight}}
\caption{Time distribution for fixed delay pulses, bin width is 1.74~ns.}
\label{f:DRPULSE}
\end{figure}

The drift chamber boards are more complicated in functionality.
Their time measurement capability was paid special attention to 
during the testing. One of the tests is briefly described below.
A random pulse derived from a PMT noise was applied to one or 
several channel inputs, while the trigger was made of this
signal by delaying it on a fixed cable line for a few hundred nanoseconds. 
The drift board essentially measures the time between the signal and 
the trigger, so one would expect constant measured time for this test.
The result of this test is presented in Fig.~\ref{f:DRPULSE}. 
The signal and trigger arrival times are digitized independently,
so it's quite natural that the measured difference is distributed
between two adjacent bins. Monotonicity
and linearity of the conversion was also checked by changing the
cable delay in 0.5~ns steps and recording the peak positions.

\begin{figure}
\centerline{\epsfig{figure=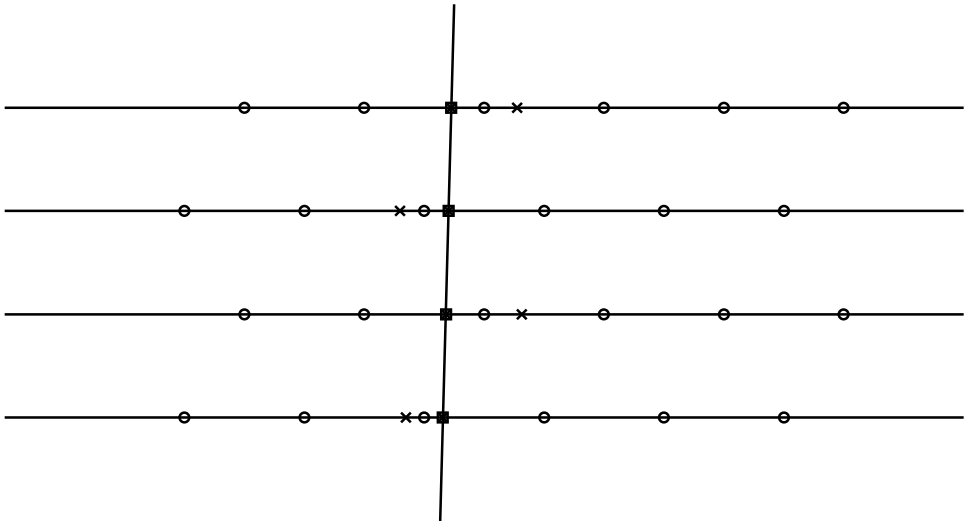, width=0.75\textwidth}}
\caption{4-hit track in the barrel; circles denote the sensitive wire positions
in 4 planes, crosses -- hits on both sides of the wire, 
squares -- hits, selected for the track, vertical line -- best~$\chi^2$ straight
track.}
\label{f:DRPICT}
\end{figure}

Most of the beam tests of the drift board were performed with a small
prototype drift chamber with hexagonal structure, which is designed
in a form of a barrel. Cell diameter is 20~mm and 24 cells are arranged
into 4 sensitive planes, two of them shifted half cell diameter. Such
design allows to perform best $\chi^2$ track reconstruction like 
it is shown in Fig.~\ref{f:DRPICT}. 

\begin{figure}[!b]
\centerline{
\begin{tabular}{cc}
\epsfig{figure=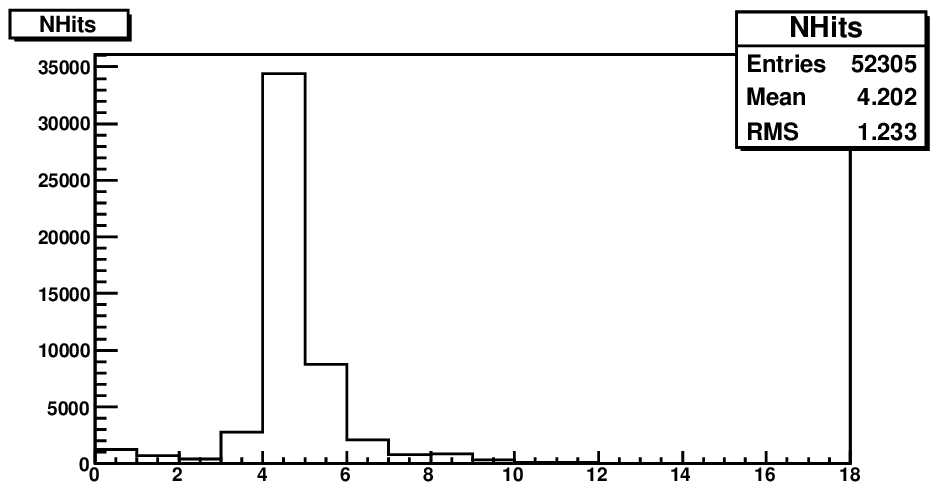, width=0.45\textwidth,
height=0.22\textheight} &
\epsfig{figure=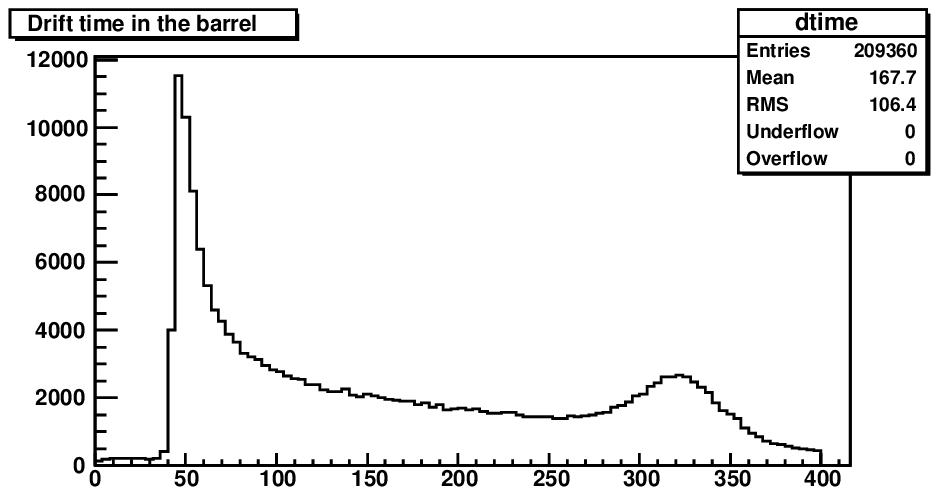, width=0.45\textwidth,
height=0.22\textheight} \\
a) & b) \\
\end{tabular}
}
\caption{Number of hits per trigger (a) and drift time distribution (b) 
in the barrel.}
\label{f:DRNHPT}
\end{figure}

Looking at Fig.~\ref{f:DRNHPT}a which presents the number of hits per trigger
in the barrel exposed to a focused beam perpendicular to the sensitive planes,
one can make conclusions on the chamber efficiency. About 67\% of events
give exactly one hit in each plane, resulting in strong maximum
at 4 hits per trigger. Small fraction of 5- and 6-hit events (about 20\%)
represents the cases, when in one or two planes charge division between
adjacent cells occurred. Number of 3-hit events is a real measure
of a single plane inefficiency and the resulting number is $(1-\varepsilon)=$4.7\%. 
Noticeable amount of zero-hit events can be accounted by the trigger
quality as well as by the effect of the trigger window overlapping, which
becomes not negligible at trigger rates as high as $10^5$~s$^{-1}$ and
1~$\mu$s window length. It's worth mentioning that the efficiency picture 
remains essentially the same in  wide ranges of applied high voltages
and discriminator threshold settings.

Figure~\ref{f:DRNHPT}b presents the distribution of the drift time
in the cells with defocused perpendicular beam (nearly flat beam
density over the cell size). Due to the $1/r$ dependence of the
drift field the distribution density significantly increases for
small drift times, while the bump on the right side of the figure
corresponds to the very long drift times of the charges, produced
in the area of zero field close to the cell boundary. This distribution
can be used for building the drift function $X(t_{DRIFT})$ giving
the hit spatial coordinate by the measured drift time.

\begin{figure}[!b]
\centerline{\epsfig{figure=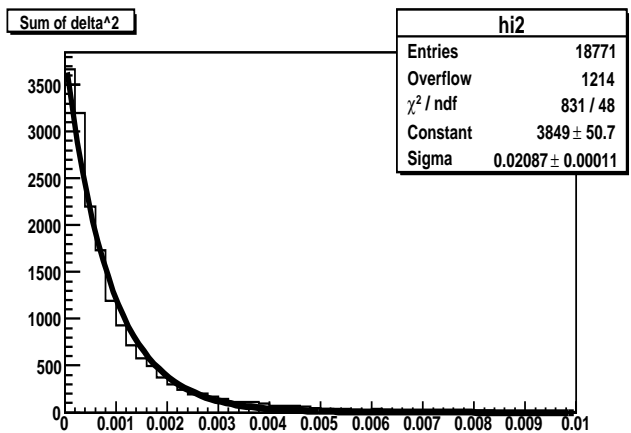, width=0.5\textwidth,
height=0.25\textheight}}
\caption{Spatial accuracy of the drift chamber prototype. Fit parameters:
'Sigma' -- the exponent value, 'Constant' -- the normalization factor.}
\label{f:DRCHI2}
\end{figure}

The final goal of the tests of the drift chamber in conjunction with
the designed electronic boards was to understand the spatial accuracy
of the whole system. The capability of the barrel for the straight
track reconstruction allows to perform the accuracy analysis. For
4-hit tracks the distribution of sums of deviations squared should
be a $\chi^2$ distribution with 2 degrees of freedom which has
a form of exponential function. If properly normalized, the exponent
value gives the average hit deviation from the best straight track,
i.e. the coordinate accuracy of a single drift cell. Such a $\chi^2$
distribution together with its exponential fit is presented
in Fig.~\ref{f:DRCHI2}. The cell radius is taken for the coordinate
measurement unit, so that the resulting exponent value should be
multiplied by 8.5~mm to get the cell accuracy (180~$\mu$m).   
The accuracy has only very weak dependence on the high voltage
on the potential wires (i.e. the gaseous amplification) and on
the discriminator threshold setting, even in cases when the
cell efficiency significantly drops. This result is not
completely understood yet, since all the empirical considerations
state that the accuracy should decrease when the signal amplitude
becomes smaller.
\section{Conclusions}

The data acquisition system for the proportional and drift
chambers of the 'EPECUR' experiment was designed, tested and produced
in ITEP in less than 2~years. Taking advantages of modern electronic
technologies a really cost effective and compact DAQ system was build
and showed excellent performance features. Tests with chamber
prototypes showed good efficiency even with 1~mm proportional
chmabers and resulted in 180~$\mu$m spatial accuracy in hexagonal
drift cells. 

The system is a keystone
of the 'EPECUR' experiment and the 'EPECUR' team hopes that the
fast progress in the setup construction in spite of all difficulties
will lead them soon to excellent physics results.

\section{Acknowledgments}

The authors express heartiest thanks to Dr. Nikolay Bondar', the member
of PNPI HEPD Radio Electronic Department, for useful discussions
on the chamber electronics.

'EPECUR' project is carried out under the financial support
of the Russian Federal Agency on Atomic Energy.
The work is also partially supported by the Russian Fund for Basic
Research grant 05-02-17005a.

\clearpage

\appendix
\renewcommand{\thetable}{\thesection\arabic{table}}
\setcounter{table}{0}
\setcounter{section}{1}

\section*{Appendix. Proportional and Drift Boards \linebreak Programming Reference}

\begin{table}[!h]
\caption{Proportional and drift boards USB endpoints and requests.}
\label{t:PRUSB}
\begin{center}
\small
\begin{tabular}{|c|c|c|c|c|c|p{0.35\textwidth}|}
\hline
\multicolumn{7}{|c|}{{\bf Endpoint 0x0/0x80}: 
Standard control endpoint$^1$ with added specific requests} \\
\hline
RqType & Request & Value & Index & Length & Data & Description \\
\hline
0x40 & 0x20 & X & X & 0 & -- & Generate PROG pulse to FPGA to start configuration \\
\hline
0x40 & 0x21 & X & X & len & data[len] & Load bitstream of 'len' bytes to FPGA \\
\hline
0x40 & 0x22 & data & 0 & 0 & -- & Write low byte of 'data' to threshold DAC \\
\hline
0x40 & 0x22 & X & 1 & 0 & -- & Enable dataflow from FPGA to USB Endpoint 0x82\\
\hline
0x40 & 0x22 & X & 2 & 0 & -- & Disable dataflow from FPGA to USB Endpoint 0x82\\
\hline
0x40 & 0x22 & X & 3 & 0 & -- & Generate a test pulse on the internal line \\
\hline
0x40 & 0x22 & X & 4 & 0 & -- & Generate a pulse on the test output \\
\hline
0x40 & 0x22 & X & 5 & 0 & -- & Send 'Packet End' to USB Endpoint 0x82 \\
\hline
0x40 & 0x23 & X & 0 & 16 & data[16] & Write 16 bytes of 'data' to I$^2$C EEPROM \\
\hline
0x40 & 0x24 & data & addr & 0 & -- & Write low byte of 'data' to FPGA register \#'addr' \\
\hline
0xC0 & \parbox[t][2.2em]{1cm}{0x20 \\ 0x21} & X & X & 1 & done & Read FPGA DONE condition \\
\hline
0xC0 & 0x22 & X & X & 4 & data[4] & Read status (data[0]), DAC setting (data[1]) 
and board temperature$^2$ (data[2]\&data[3], 10 bit left justified)\\
\hline
0xC0 & 0x23 & X & 0 & 16 & data[16] & Read 16 bytes from I$^2$C EEPROM \\
\hline
0xC0 & 0x24 & X & addr & 1 & data & Read byte from FPGA register \#'addr' \\
\hline
\multicolumn{7}{|c|}{{\bf Endpoint 0x82}: 
Bulk endpoint (only IN direction)} \\
\hline
\multicolumn{7}{|c|}{Main data endpoint, for the dataflow structure  see Table~\ref{t:PRFMT}} \\
\hline
\end{tabular}
\end{center}
\vspace{-0.5em}
{\bf Notes:}
\vspace{-0.8em}
\begin{enumerate}
\itemsep 0cm
\parskip 0cm
\item For USB standard and specific to CYPRESS $\mu$processor Endpoint 0x0/0x80 
requests see~\cite{c:USB} and~\cite{c:CYPRESS}.
\item Board temperature fields are only meaningful for the drift chamber board.
\end{enumerate}
\end{table}

\begin{table}
\caption{Proportional board dataflow structure.}
\label{t:PRFMT}
\begin{center}
\small
\begin{tabular}{|c|c|c|c|p{0.55\textwidth}|}
\hline
Byte\# & Bit7 & Bit6 & Bits[5:0] & Description \\ 
\hline
0 & 1 & 0 & TRNUM$_1$[12:7] & Trigger signature and upper part of trigger number \\
\hline
1 & 0 & \multicolumn{2}{|c|}{TRNUM$_1$[6:0]} & Lower part of trigger number \\
\hline
2 & 0 & \multicolumn{2}{|c|}{CHNUM$_1$[6:0]} & Number of first channel hit in this trigger \\
\hline
\dots & 0 & \multicolumn{2}{|c|}{\dots} & Numbers of other channels hit in this trigger \\
\hline
n+2 & 0 & \multicolumn{2}{|c|}{CHNUM$_n$[6:0]} & Number of last channel hit in this trigger \\
\hline
n+3 & 1 & 0 & TRNUM$_2$[12:7] & Trigger signature and upper part of next trigger number \\
\hline
n+4 & 0 & \multicolumn{2}{|c|}{TRNUM$_2$[6:0]} & Lower part of next trigger number \\
\hline
n+5 & 0 & \multicolumn{2}{|c|}{CHNUM[6:0]} & Number of first channel hit in this trigger \\
\hline
\dots & 0 & \multicolumn{2}{|c|}{\dots} & Numbers of other channels hit in this trigger \\
\hline
\dots & \dots & \multicolumn{2}{|c|}{\dots} & Other trigger blocks in this cycle \\
\hline
m-1 & 1 & 1 & CYNUM[12:7] & Cycle signature and upper part of cycle number \\
\hline
m & 0 & \multicolumn{2}{|c|}{CYNUM[6:0]} & Lower part of cycle number \\
\hline
\dots & \dots & \multicolumn{2}{|c|}{\dots} & Next cycle data \\
\hline
\end{tabular}
\end{center}
\end{table}

\begin{table}
\caption{Drift board dataflow structure.}
\label{t:DRFMT}
\begin{center}
\small
\begin{tabular}{|c|c|c|c|p{0.5\textwidth}|}
\hline
Byte\# & Bit7 & Bit6 & Bits[5:0] & Description \\ 
\hline
0 & 1 & 1 & 111111$_b$ & Trigger signature  \\
\hline
1 & 0 & 0 & TRNUM$_1$[5:0] & Lower part of trigger number \\
\hline
2 & 0 & 0 & TRNUM$_1$[11:6] & Upper part of trigger number \\
\hline
3 & 0 & 1 & 0 \& CHNUM$_1$[4:0] & First hit channel number \\
\hline
4 & 0 & 0 & TIME$_1$[5:0] & Lower part of this channel time \\
\hline
5 & 0 & 0 & TIME$_1$[11:6] & Upper part of this channel time \\
\hline
\dots & 0 & \multicolumn{2}{|c|}{\dots} & Numbers and times of other channels hit in this trigger \\
\hline
3n & 0 & 1 & 0 \& CHNUM$_n$[4:0] & Last hit channel number \\
\hline
3n+1 & 0 & 0 & TIME$_n$[5:0] & Lower part of this channel time \\
\hline
3n+2 & 0 & 0 & TIME$_n$[11:6] & Upper part of this channel time \\
\hline
3n+3 & 1 & 1 & 111111$_b$ & Next trigger signature  \\
\hline
3n+4 & 0 & 0 & TRNUM$_2$[5:0] & Lower part of next trigger number \\
\hline
3n+5 & 0 & 0 & TRNUM$_2$[11:6] & Upper part of next trigger number \\
\hline
\dots & 0 & \multicolumn{2}{|c|}{\dots} & Numbers and times of channels hit in this trigger \\
\hline
\dots & \dots & \multicolumn{2}{|c|}{\dots} & Other trigger blocks in this cycle \\
\hline
3k & 1 & 0 & 111111$_b$ & Cycle signature \\
\hline
3k+1 & 0 & 0 & CYNUM[5:0] & Lower part of cycle number \\
\hline
3k+2 & 0 & 0 & CYNUM[11:6] & Upper part of cycle number \\
\hline
\dots & \dots & \multicolumn{2}{|c|}{\dots} & Next cycle data \\
\hline
\end{tabular}
\end{center}
\end{table}

\begin{table}
\caption{Proportional board FPGA 8 bit registers.}
\label{t:PRXREG}
\begin{center}
\small
\begin{tabular}{|p{0.07\textwidth}|l|p{0.75\textwidth}|}
\hline
Address & Name & Fields and description \\
\hline
0 RW & CSR0 & CSR0[7:0]=DLY[7:0] -- The delay of the 100-bit digital
delay line, CLK$^1$ periods \\
\hline
1 RW & CSR1 & CSR1[0]=DLY[8] \par
CSR1[3:1]=DTIM[2:0] -- Channel dead time: $T_{DEAD}=2+DTIM*9$ CLK periods \par
CSR1[5:4]=OLEN[1:0] -- Channel output duration for putting to the digital 
delay line, $2^{OLEN}$ CLK periods \par
CSR1[6] -- Trigger source select: 0 -- External from RJ45 connector; 1 -- Internal,
grand OR of all channels OR generated by SFT\_TR \par
CSR1[7] -- Enable data transfer to the $\mu$processor \\
\hline
2 RW & CSR2 & CSR2[7:0]=TRGDLY[7:0] -- Additional delay of the grand OR trigger
to the output, CLK periods; if TRGDLY=0, output is forbidden \\
\hline
3 RW & CSR3 & CSR3[3:0]=TRGLEN[3:0] -- Length of the grand OR trigger on the
output, TRGLEN+1 CLK periods \par
CSR3[6:4] -- Reserved \par
CSR3[7] -- Enable accelerator cycle emulator (period 2.8~s, inhibit 0.35~s) \\
\hline
4 W & RST & Various resets, selected by the written mask: \par
RST[0] -- Reset all FIFOs \par
RST[1] -- Trigger counter clear \par
RST[2] -- Cycle counter clear \par
RST[7:3] -- Reserved \\
\hline
5 W & SFT\_TR & Writing of any number causes a trigger if enabled
with CSR1[6] \\
\hline
4--5 R & CYNUM & CYNUM[15:0] -- Last cycle number \\
\hline
6--18 RW & ORMSK & ORMSK[99:0]$^2$ -- Bit mask inhibiting individual channel
participation in the grand OR trigger output;  \\
\hline
19--31 RW & MASK & MASK[99:0]$^2$ -- Bit mask completely inhibiting individual
channels \\
\hline
32--44 R & INLVL & INLVL[99:0]$^3$ -- Bit representation of logic levels on
individual channel inputs \\
\hline
45 R & 0 & Always read 0 \\
\hline
46--47 R & TRNUM & TRNUM[15:0] -- Last trigger number \\
\hline
\end{tabular}
\end{center}

{\bf Notes:}
\begin{enumerate}
\itemsep 0cm
\parskip 0cm
\item CLK frequency is 96~MHz.
\item 4 most significant bits are RW, but have no effect.
\item 4 most significant bits always read 0
\end{enumerate}
\end{table}

\begin{table}
\caption{Drift board FPGA 8 bit registers.}
\label{t:DRXREG}
\begin{center}
\small
\begin{tabular}{|p{0.09\textwidth}|l|p{0.72\textwidth}|}
\hline
Address & Name & Fields and description \\
\hline
0 RW & CSR0 & CSR0[7:0]=WINBEG[7:0] -- Trigger window start prior
to the trigger, lower part, CLK1$^1$ periods \\
\hline
1 RW & CSR1 & CSR1[2:0]=WINBEG[10:8] -- Trigger window start, upper part \par
CSR1[7:3]=WINLEN[4:0] -- Trigger window length$^2$, lower part, CLK1$^1$ periods \\
\hline
2 RW & CSR2 & CSR2[4:0]=WINLEN[9:5] -- Trigger window length, upper part \par
CSR2[7:5]=DTIM[2:0] -- Channel dead time: $T_{DEAD}=2+DTIM*5$ CLK2$^3$ periods \\
\hline
3 RW & CSR3 & CSR3[0] -- Trigger source select: 0 -- External from RJ45 connector;
1 -- Internal, from the $\mu$processor \par 
CSR3[2:1] -- Power monitor select: \parbox[t][4.3em][t]{0.35\textwidth}
    {\begin{itemize}
     \itemsep 0cm
     \parskip 0cm
     \itemindent -1cm
     \vspace{-1.7em}
     \item[] 0 - positive analog supply
     \item[] 1 - comparator digital supply
     \item[] 2 - FPGA core supply
     \item[] 3 - negative analog supply
     \end{itemize}} \par
CSR3[6:3] -- Reserved \par
CSR3[7] -- Enable data transfer to the $\mu$processor \\
\hline
4 W & RST & Various resets, selected by the written mask: \par
RST[0] -- Reset all FIFOs, emulate cycle end \par
RST[1] -- Trigger counter clear \par
RST[2] -- Cycle counter clear \par
RST[7:3] -- Reserved \\
\hline
5--7 RW & MASK & MASK[23:0] -- Bit mask completely inhibiting individual
channels \\
\hline
8--10 R & INLVL & INLVL[23:0] -- Bit representation of logic levels on
individual channel inputs \\
\hline
11 R & PWMON & Power monitor reading, selected by CSR3[2:1] \\
\hline
12--13 R & CYNUM & CYNUM[15:0] -- Last cycle number \\
\hline
14--15 R & TRNUM & TRNUM[15:0] -- Last trigger number\\
\hline
\end{tabular}
\end{center}

{\bf Notes:}
\begin{enumerate}
\itemsep 0cm
\parskip 0cm
\item CLK1 frequency is 144~MHz.
\item The trigger window ends not later than the actual trigger comes,
so that values greater than WINBEG are useless.
\item CLK2 frequency is 48~MHz.
\end{enumerate}
\end{table}

\begin{table}
\caption{Proportional and drift board LED indicators.}
\label{t:PRLED}
\begin{center}
\small
\begin{tabular}{|c|c|c|p{0.7\textwidth}|}
\hline
Prop. & Drift & Color & Description \\
\hline
D53 & D2 & Green & +5V Analog power \\
\hline
D54 & D3 & Green & -5V Analog power \\
\hline
D60 & D4 & Green & +3.3V Digital power \\
\hline
D56 & D30 & Yellow & Heartbeating: regularly flashes when the application
code is running in the $\mu$processor \\
\hline
D57 & D31 & Yellow & Flashes on USB control endpoint requests\\
\hline
D58 & D32 & Yellow & Flashes on a trigger from any source accepted by the board, 
lights between accelerator cycles \\
\hline
D59 & D33 & Yellow & Flashes on data transfers from FPGA to the $\mu$processor \\
\hline
D61 & D1 & Red & Lights when the FPGA is not configured \\
\hline
\end{tabular}
\end{center}
\end{table}

\begin{table}
\caption{Proportional and drift board power and trigger connector pinout.}
\label{t:PRCONN}
\begin{center}
\small
\begin{tabular}{|c|p{0.5\textwidth}|}
\hline
Pin\# & Description \\
\hline
1 & LVDS trigger input \\
\hline
2 & LVDS trigger input complimentary \\
\hline
3 & Power +6~V \\
\hline
4 & Power +4~V \\
\hline
5 & Power $-6$~V \\
\hline
6 & Power return ground \\
\hline
7 & LVDS trigger output$^1$ \\
\hline
8 & LVDS trigger output complimentary$^1$ \\
\hline
\end{tabular}
\end{center}

{\bf Notes:}
\begin{enumerate}
\itemsep 0cm
\parskip 0cm
\item In drift chamber boards the LVDS trigger outputs are
always driven passive.
\item A standard UTP Ethernet cable can be used for the power
and trigger connection.
\end{enumerate}
\end{table}

\clearpage
\thispagestyle{empty}

%\clearpage
%\thispagestyle{empty}
%\rule{\textwidth}{0cm}
\vfill
\noindent\rule{\textwidth}{0.75pt}
\vspace{0.3em}
\foreignlanguage{russian}{\parbox{\textwidth}{
\hspace{0.4cm}Подписано к печати 14.12.06.\hfillФормат~60х90\hfill1/16\hspace{0.4cm} }}
\foreignlanguage{russian}{\parbox{\textwidth}{
\hspace{0.4cm}Усл.-печ.~л.~1,5\hfill Уч.-изд.~л.~1,1\hfill Тираж 120~экз.\hfill Заказ~21\hspace{0.4cm} }}
\foreignlanguage{russian}{\parbox{\textwidth}{\hfilИндекс 3649\hfill }}
%\vspace{-1em}
\noindent\rule{\textwidth}{0.75pt}
\foreignlanguage{russian}{\parbox{\textwidth}{
\hfilОтпечатано в ИТЭФ, 117218, Москва, ул.~Б.~Черемушкинская, д.~25\hfil}}

\clearpage
\thispagestyle{empty}
\noindent
\foreignlanguage{russian}{\parbox{\textwidth}{\Large \bf \hfill Индекс 3649}}
\vfill
\noindent
\foreignlanguage{russian}{\parbox{\textwidth}{\Large \bf \hfil Препринт 21--06, ИТЭФ, 2006\hfill}}


\begin{thebibliography}{9}
\itemsep 0mm
\parskip 0mm
\bibitem{c:EPECUR}
I.G.~Alekseev et al. Search for the Cryptoexotic Member of the Baryon
Antidecuplet 1/2$^+$ in the Reactions $\pi^-p\to \pi^-p$ and 
$\pi^-p\to K\Lambda$. Preprint ITEP 2-05; {\tt \bf hep-ex/0509032}.
\bibitem{c:KATZ}
L.Z.~Barabash et al. Methods and Results of Tuning the Universal 
Secondary Particle Channel of the Reconstructed ITEP Proton Synchrotron.
Preprint ITEP 115-1975.
\bibitem{c:SACLAY}
G.D.~Alkhazov et al. Forward Spectrometer for Study of Low-lying
Baryon Resonances. SPES4-$\pi$ Experiment at Saturne-II Accelerator (Saclay).
Preprint PNPI 2352-EP-9-2000.
\bibitem{c:XILINX}
Spartan-3 Complete Data Sheet. {\tt http://direct.xilinx.com/bvdocs/} 
{\tt publications/ds099.pdf}.
\bibitem{c:CYPRESS}
EZ-USB FX2LP$^{\rm TM}$ USB Microcontroller. {\tt http://download.cypress.com/} 
{\tt published-content/publish/design\_resources/datasheets/con-}
{\tt tents/cy7c68013a\_8.pdf}.
\bibitem{c:USB}
Universal Serial Bus Revision 2.0 Specification. \\
{\tt http://www.usb.org/developers/docs/usb\_20\_05122006.zip}.
\bibitem{c:CPLD}
XC2C32A CoolRunner-II CPLD. \\
{\tt http://direct.xilinx.com/bvdocs/publications/ds310.pdf}.
\end{thebibliography}
\end{document}